\documentclass[a4paper,twocolumn]{revtex4}

\usepackage{latexsym}
\usepackage{graphics}
\usepackage{color}
\usepackage[latin1]{inputenc}
\usepackage{amssymb}
\usepackage{amsmath}
\usepackage{amsfonts}
\usepackage[dvips]{graphicx}

\usepackage{pdfpages}
\usepackage{epstopdf}

\begin{document}

\title{Effective velocity distribution in an atom gravimeter: effect of the convolution with the response of the detection}

\author{T. Farah, P. Gillot, B. Cheng, A. Landragin, S. Merlet, F. Pereira dos Santos}                  

 \affiliation{LNE-SYRTE, Observatoire de Paris, LNE, CNRS, UPMC \\
61 avenue de l'Observatoire, 75014 Paris, France}

\begin{abstract} We present here a detailed study of the influence of the transverse motion of the atoms in a free-fall gravimeter. By implementing Raman selection in the horizontal directions at the beginning of the atoms free fall, we characterize the effective velocity distribution, ie the velocity distribution of the detected atom, as a function of the laser cooling and trapping parameters. In particular, we show that the response of the detection induces a pronounced asymetry of this effective velocity distribution that depends not only on the imbalance between molasses beams but also on the initial position of the displaced atomic sample. This convolution with the detection has a strong influence on the averaging of the bias due to Coriolis acceleration. The present study allows a fairly good understanding of results previously published in {\it Louchet-Chauvet et al., NJP 13, 065025 (2011)}, where the mean phase shift due to Coriolis acceleration was found to have a sign different from expected. 

\end{abstract}

\maketitle

\section{Introduction}

Atom gravimeters compete favourably with state of the art corner-cube gravimeters, both in terms of sensitivity and accuracy. Their sensitivity can reach a level as low as $4.2 \mu$Gal at 1 s measurement time \cite{Hu2013}, and their relative accuracy is of a few $10^{-9}$g \cite{Peters2001,Louchet-Chauvet2011}. Direct comparisons between the two technologies, with several such instruments performing measurements at the same place and same time, have shown the capability of atomic devices to reach better sensitivities than their optical counterparts \cite{Peters2001,Merlet2010,Gillot2014}. In particular, their higher repetition rate, up to several Hz, allows for a better filtering of low frequency ground vibration noise. In addition, the parasitic vibration noise is significantly reduced thanks to efficient antivibration systems, based on active or passive solutions, and/or correlations with auxiliary motion sensors, such as sismometers or accelerometers. Remarkably, the latter solution allows the atomic devices to operation of the device in the presence of large levels of vibration noise, on the ground \cite{Merlet2009} or in an airplane \cite{Geiger2011}. As for their accuracy, atom gravimeters are limited by effects related to the motion of the atoms in the laser beams used to separate and recombine the atomic wavepackets in the interferometer, namely Coriolis acceleration and effects related to wavefront aberrations. The first effect has been identified and studied in the early experiment of \cite{Peters2001}, where a rotation of the whole experimental setup was used to modulate, and eventually cancel, the effect of the Earth rotation rate. Alternatively, a synchronous rotation of the single mirror that is used to retroreflect the interferometer lasers has been used \cite{Lan2012,Dickerson2013}. This offers the possibility to cancel Coriolis shifts and thus increase the contrast of the interferometer in the case where the latter is reduced by the dispersion of the Coriolis shifts due to the velocity spread of the atomic sample. This compensation was used in differential interferometers \cite{Lan2012,Sorrentino2014}, as well as in "single" interferometers, such as gravimeters \cite{Dickerson2013,Hauth2013}. In the latter case, a careful control of the mirror motion is necessary to avoid any residual synchronous vertical acceleration, that would bias the gravity measurement. Remarkably, the use of large free fall times and ultracold samples offers the possibility to map transverse effects by measuring the populations in the two output ports of the interferometer with a spatially resolved CCD imaging, as demonstrated in \cite{Sugarbaker2013}. In this case, one measures not only the average value, but also the dispersion of the phase shifts due to Earth rotation and wavefront aberrations, which in principle could also help to reconstruct the wavefront of the lasers, and make a precise determination of its influence in the measurement. Another method to separate the shift due to Coriolis acceleration from other systematic effects is to perform measurements for two opposite orientations of the experiment in the horizontal plane, which inverts the orientations of the atomic velocities with respect to the Earth rotation vector, and thus changes the sign of the Coriolis acceleration. Calculating the half difference between these two measurements, one gets the shift due to Coriolis acceleration and averaging the two measurements, the interferometer phase is corrected from Coriolis acceleration. We use this last method in \cite{Louchet-Chauvet2011}, as well as during comparison campaigns \cite{Merlet2009,Jiang2012,Francis2013}. 

The paper provides a quantitative understanding of the amplitude of the Coriolis shift measured in \cite{Louchet-Chauvet2011} and its dependance on the experimental parameters. We start by recalling the main features of our cold atom gravimeter, and then present a detailed study of the influence of the power imbalance between the trapping beams onto the measurement of gravity, as a function of the molasses detuning and overall laser intensity. The results of this study highligth the role played by the detection in the averaging of the Coriolis shift, which we confirm by shifting the initial position (and thus its position in the detection beams) of the atom cloud. We then implement Raman selection in the horizontal directions, that allows for a precise determination of the velocity distribution of the detected atoms, and a quantitative analysis of the influence of the convolution with the response of the detection. We finally perform gravity measurements with atoms selected in the horizontal direction. In particular, we measure the Coriolis shift as a function of the selected velocity, and show that the Coriolis bias can be corrected for by selecting atoms in a well centered and narrow velocity distribution.

\section{Atom interferometer}

We perform the experiments in an atom gravimeter that has been previously described in detail in \cite{Louchet-Chauvet2011}. We recall here its main features. We first trap from a 2D MOT about $10^7$ atoms in a three dimensional magneto-optical trap (MOT) within 70 ms. The quadrupole field of the MOT is then switched off and the cooling laser is first tuned for 8 ms to a different detuning (ranging from $-1.9\Gamma$ to $-6.6\Gamma$ depending on the measurements), and then to $-20\Gamma$ within 1 ms. A few ms long far off detuned molasses phase, followed by an adiabatic extinction of the lasers, cools the atoms to about $2~\mu$K. After their release from the trap, the atoms are in the $F=2$ state. We then apply a 10 mG bias field in order to lift the degeneracy between Zeeman sensitive transitions. The atoms are then selected in a narrow velocity distribution ($\delta v \sim v_r$) in the $|F=1,m_F=0\rangle$ state using a combination of several microwave pulses resonant with the $|F=2,m_F=0\rangle \rightarrow |F=1,m_F=0\rangle$ transition, pushing laser pulses and a Raman vertical selection pulse. The atoms then interact with vertical counterpropagating Raman lasers, in a sequence of three $\pi/2-\pi-\pi/2$ pulses, realizing a Mach Zehnder type interferometer. The duration of the $\pi/2$ (resp. $\pi$) pulse is typically of order of 10 (resp. 20) $\mu$s. The total interferometer time is 140 ms. After the interferometer, the atoms are detected by a state selective fluorescence detection that allows measurement of the two populations in the two output ports \cite{Borde1989}. From this population measurement, we derive the phase of the interferometer. The total measurement cycle time is 360 ms.

\section{Influence of the molasses parameters}


We begin performing a study of the influence of the intensity imbalance of the trapping beams onto the measurement of gravity. This intensity imbalance is given by $x_{ij}=(I_j-I_i)/(I_i+I_j)$, where $I_i$ and $I_j$ are the intensities in two counterpropagating beams $i$ and $j$. Beams $5$ and $6$ propagate along the East-West, while the other four MOT beams, ($1,2$ and $3,4$) are propagating in the North-South vertical plane, and are tilted by $45^{\circ}$ with respect to the horizontal plane (see figure \ref{fig:setup}). The intensities in all six MOT beams are measured thanks to photodiodes installed behind $45^{\circ}$ mirrors installed in the (angled) MOT beams collimators. Imbalancing the intensities in the trapping beams has two effects : first, it induces a drift velocity during the molasses phases, and thus a non-zero mean transverse velocity after the release from the far-detuned molasses. Second, it changes the initial position of the atomic sample, because 1) the position of the MOT gets shifted, 2) the atoms drift away from the MOT's position during the molasses phase(s). Changes on the position and velocity of the atoms will affect the phase of the interferometer through the two transverse effects discussed above, that are difficult to separate : Coriolis acceleration, if the atoms have a non zero transverse velocity along the EW direction, and wavefront distortions, which depends on the position/trajectories of the atoms in the Raman laser beams.

\begin{figure}[ht]
        \centering
        \includegraphics[width=8 cm]{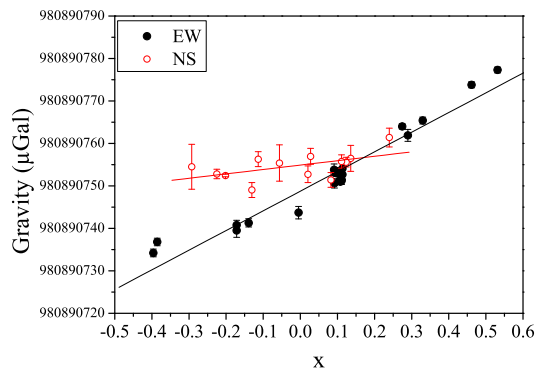}
\caption{Gravity measurement as a function of the intensity unbalance between molasses beams along the EW and NS directions. The detuning of the first molasses phase is $-1.9\Gamma$.}
        \label{fig:gVsx}
   \end{figure}

\begin{figure}[ht]
        \centering
        \includegraphics[width=8 cm]{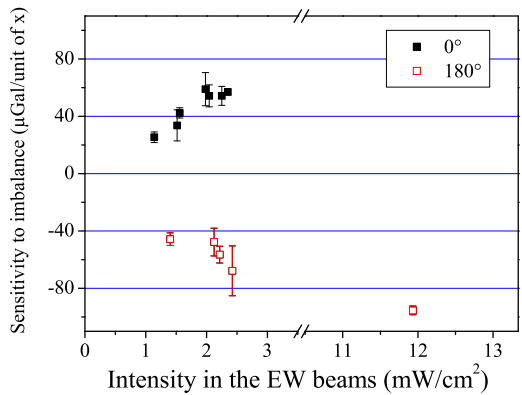}
\caption{Sensitivity to the intensity unbalance as a function of the total laser intensity in the EW beams, for two opposite orientation of the experiment. The detuning during the first molasses phase is $-1.9\Gamma$}
        \label{fig:gVsxAll}
   \end{figure}

Figure \ref{fig:gVsx} displays the results of the measurement of gravity acceleration in $\mu$Gal ($1~\mu$Gal=$10^{-8}$m/s$^2$) as a function of the power imbalance along each direction. For that measurement, the total intensity in the EW beams is 1.6 mW/cm$^2$ and the detuning of the first molasses phase is $-1.9\Gamma$. We observe a larger effect along the EW direction, with a sensitivity of 46(2) $\mu$Gal/unit of $x_{56}$, than along the NS (10(5) $\mu$Gal/unit of $x_{NS}$). The NS measurement was realized by unbalancing one of the other pairs of MOT beams. This tends to indicate that changes of the bias due to Coriolis acceleration, which affect only the EW and not the NS direction, are larger than the changes in the aberration shift. This is confirmed by performing the same measurements after having rotated the experiment by $180^{\circ}$. As shown previously in \cite{Louchet-Chauvet2011}, the effect along the EW direction is of the same order of magnitude, with an opposite sign. Figure \ref{fig:gVsxAll} displays the results of the measurements of the sensitivity to imbalance for different laser powers in the EW beams and two opposite orientations of the experiment. We indeed observe the change in the sign of the sensitivity to imbalance when rotating the experiment by $180^{\circ}$, and an increase of this sensitivity with increasing laser cooling power in the EW beams. 

As explained earlier, averaging the measurements with the two opposite orientations allowed us to correct for the Coriolis shift. Yet, the sign of the effect is found to be different from expected : inducing an imbalance with less intensity in the East beam induces a net mean velocity oriented towards East and thus a negative Coriolis aceleration bias, while we observe a positive bias on the gravity measurement. This can be explained by a clipping effect in the detection.

\section{Response of the detection}
\begin{figure}[ht]
        \centering
        \includegraphics[angle=270,width=8 cm]{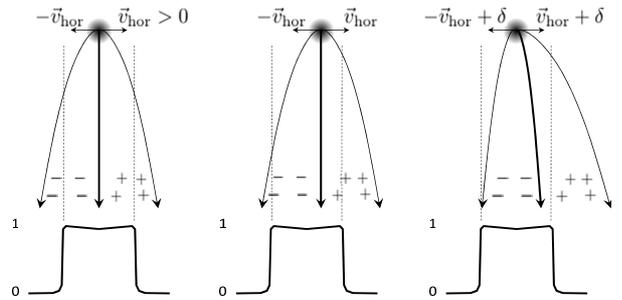}
\caption{Effect of the convolution by the detection system. The normalized response function of the detection along the EW direction, displayed at the bottom, is close to a square shape function (see \cite{Louchet-Chauvet2011} for more details). Left: Case of a cloud well centered with respect to the detection. The atoms with high initial transverse velocity fall outside the field of the detection system. Middle: Case of an off-centered cloud. When the initial position is shifted to the right with respect to the center of the detection, atoms with large velocities to the right are less efficiently detected. This results in an asymetry of the effective velocity distribution. Right: When the initial mean velocity of the cloud $v_0$ is not null, the clipping by the detection results in an average detected velocity lower than $v_0$.}
        \label{fig:clipping}
   \end{figure}

The effect of the clipping due to the detection is illustrated in figure \ref{fig:clipping}. The field of view of the detection system is limited to 14 mm in the EW detection, due to the finite size of the photodiodes that collect the fuorescence emitted by the atoms (see \cite{Louchet-Chauvet2011} for a detailed description of the detection system). Atoms with lie sufficiently far in the wings of the velocity distribution are not detected. If the initial position of the cloud is not centered with respect to the detection, the response of the detection induces an asymetric response when averaging over the velocity distribution. In particular, atoms with large velocities in the direction of the initial position shift are less efficiently detected, which results in a net Coriolis bias which corresponds to a mean velocity directed into the opposite direction.

\begin{figure}[ht]
        \centering
        \includegraphics[width=8 cm]{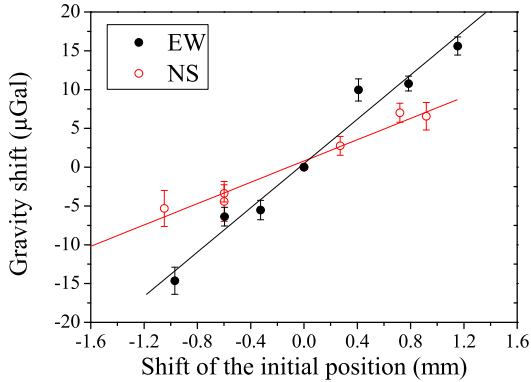}
\caption{Gravity shift as a function of the shift of the atoms from the initial position, for displacements in the EW and NS directions.}
        \label{fig:gVsBob}
   \end{figure}

To evaluate the influence of this clipping on the averaging of the Coriolis acceleration, we performed interferometer measurements with atomic samples deliberatly displaced with respect to their initial position. The displacement is induced by shifting the position of the zero magnetic field of the MOT using additional coils, while keeping the molasses beams well balanced. Two sets of Helmoltz coils were wounded on the experiment: two along the EW direction, two along the NS direction, that allow to control the position of the cloud in the horizontal plane (see figure \ref{fig:setup}). The position of the cloud right after the release from the molasses beams is measured with a CCD camera installed at $45^{\circ}$ with respect to the EW/NS directions in the horizontal plane. Figure \ref{fig:gVsBob} displays the shifts measured along the two directions. We find linear behaviours with 14.2(1.1) $\mu$Gal/mm along EW and 6.9(0.5) $\mu$Gal/mm along NS. To evaluate the impact of the displacement effect onto the measurement with unbalanced beams along the EW direction, we measured the positions (at the end of the molasses phase) as a function of the power unbalance, and found 4.95(15) mm/unit of $x_{56}$. The contribution of the displacement effect is then estimated to be 70(5) $\mu$Gal/unit of $x_{56}$. From that study we deduce that the displacement effect dominates over the effect on the mean velocity, and largely overcompensates it.    

\begin{figure}[ht]
        \centering
        \includegraphics[width=8 cm]{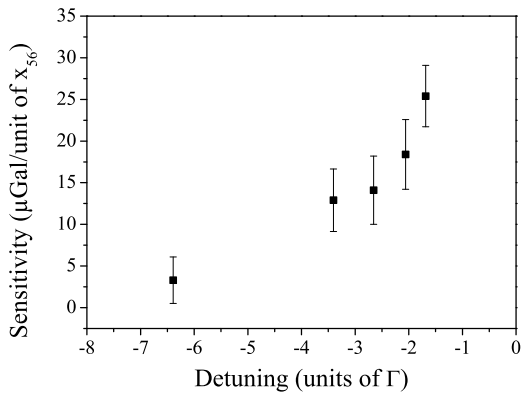}
\caption{Sensitivity to imbalance between the molasses beams in the EW direction as a function of the detuning between the first molasses phase. For that measurement, the cooling laser intensity in the EW beams is 1.1 mW/cm$^{2}$.}
        \label{fig:x56slopes}
   \end{figure}

This displacement effect can in principle be varied by changing the molasses parameters. For the measurements presented above, the detuning of the first molasses phase was set rather close to resonance, at $\sim-1.9\Gamma$, which maximises the number of detected atoms. We performed measurements of the position of the atoms immediatly after their release from the molasses as a function of the first molasses detuning, and found smaller displacements for larger detunings. In particular, operating at $\sim-6.6\Gamma$ reduces the displacement by about a factor 2, of 2.67(7) mm/unit of $x_{56}$. At this detuning the displacement is dominated by the displacement in the MOT phase, and the subsequent drift during the first molasses phases is largely reduced. We repeated the above measurements as a function of the unbalance for different detunings, and the results are displayed in figure \ref{fig:x56slopes}. We observe a significant decrease of the sensitivity to power unbalance when increasing the detuning down to $\sim-6.6\Gamma$, where the influence is reduced by a factor of about 10. At this large detuning, the effect of the clipping due to the displacement approximately compensates the effect on the mean velocity of the real velocity distribution (a drift velocity aligned in the direction of the most powerful beam). This is thus an interesting operating point for the gravimeter, as for typical fluctuations of the relative intensities in the molasses beams of order of a few \% over a few days, we expect negligible fluctuations on the gravity measurement. An even larger detuning could reduce this sensitivity even further, at the expense of a reduced number of atoms. 

In \cite{Louchet-Chauvet2011}, a calculation of the sensitivity to displacement along the EW direction was performed, considering an initial temperature of the atoms of $2 \mu$K, and taking into account the geometry of the detection. The expected sensitivity was found to be $3\mu Gal$/mm, which is much smaller than what we measure here. In \cite{Louchet-Chauvet2011}, direct gravity measurements as a function of the initial position had not been performed, but attributing all the observed shift (of about $50\mu$Gal/unit of x) to the shift of the initial position (of about 5 mm/unit of x) lead to a sensitivity of about $10\mu$Gal/mm, much smaller than the calculated sensitivity. We then assumed that asymetries in the velocity distribution when unbalancing the beams, more than position dependence, could explain the behaviour we observed, namely an inversion of the slope. From the position mesurement performed here (where the initial velocity distribution isn't affected by the displacement), we know that this assumption is not entirely valid: the position dependance is indeed much larger than expected from the calculation. To resolve the discrepancy between measurements and the calculation, a precise measurement of the velocity distribution is required. This motivated the detailed study of the horizontal velocity distribution using Raman transverse velocity selection.

\section{Horizontal selection}

\begin{figure}[h!]
        \centering
  \includegraphics[width=8 cm]{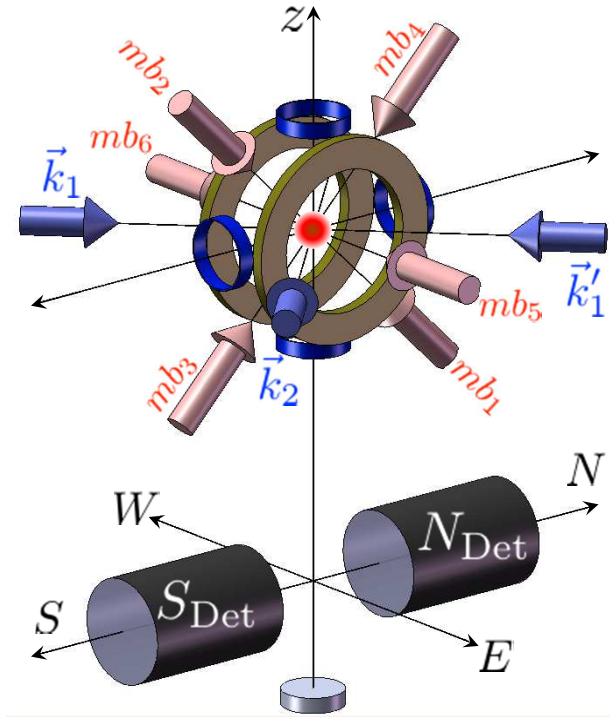}

\caption{Scheme of the MOT, horizontal Raman beams and detection setup. Two MOT beams (mb$_5$ and mb$_6$) are aligned along the EW direction. The other four are in the NS-z plane at 45$^{\circ}$ incidence with repect to the horizontal plane. The MOT coils are placed along the EW direction. Two additional pairs of coils (blue) allow for displacements of the cloud position along the z and NS directions. A last pair is wound on top of the MOT coils. Three optical access at 45$^{\circ}$ in the horizontal plane allow for velocity selecting the atoms along the two orthogonal directions EW and NS. The fourth optical access at 45$^{\circ}$ is used for imaging the cloud onto a CCD camera (not represented on the picture). The detection systems ($S_{Det}$ and $N_{Det}$) that collect the fluorescence are aligned along the NS direction. Below, the mirror used to retroreflect the vertical Raman beams is represented.}
        \label{fig:setup}
   \end{figure}

We have thus setup horizontal Raman beams, by deriving part of the power of the lasers used for the vertical Raman beams on the optical setup. We direct about 10\% of the available power of each laser, before they get mixed, in two fibers that transport these beams onto an additional small breadboard, placed above the optical bench \cite{Cheinet2006}. There, each beam is collimated and sent to an independent AOM before being recoupled in a fiber. These AOMs allow pulsing and controlling the intensity in the beams. Each fiber is then connected to a home-made collimator that enlarges the beams to a waist of 12 mm. We have installed three such collimators in front of the windows on the chamber that provide horizontal access at $45^{\circ}$ with respect to the East West direction. This allow us to perform a velocity selection either in the East-West (EW) or in the North-South (NS) direction by simply displacing one of the two fiber outputs to the third collimator. The setup is illustrated in figure \ref{fig:setup}.

\section{Effective velocity distribution}

\begin{figure}[ht]
        \centering
       \includegraphics[width=8 cm]{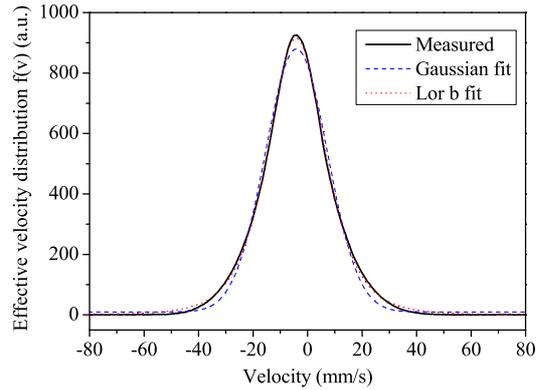}
\caption{EW effective velocity distribution. Thick back line: data, dashed line: gaussian fit, dotted line: lorentzian b fit.}
        \label{fig:distrib}
   \end{figure}

The velocity distribution in the horizontal direction can then be measured using the velocity selectivity of sufficiently long pulses of the horizontal Raman beams. Our measurement protocol, which consists in a combination of pulses instead of only one, is described in the annex, and allows for a measurement of the effective velocity distribution, without additional alteration from the response of the detection or from the Raman selection itself. Figure \ref{fig:distrib} displays as a thick continuous line the measurement of the effective velocity distribution along the EW direction performed with balanced molasses beams. 

A gaussian fit of the distribution (displayed as a dashed line) give an effective temperature of $1.4 \mu$K but clearly underestimates the wings of the velocity distribution. A fit with a Lorentzian b, $f(v)=A/(1+(v-v_0)^2/v_c^2)^b$ displayed as a thin continuous line, gives a much better agreement. We find $b=2.5$ and $v_c=22.0$. This velocity distribution has previously been introduced in the context of atomic fountains \cite{Sortais2000}, and allowed to find agreement between measured and calculated fraction of detected atoms, in the presence of several physical diaphragms and convolution of the detection. This distribution has a larger fraction of atoms in the wings, which should amplify the effect of the clipping of the detection. This is confirmed by the measurement of the mean velocity of the distribution as a function of the initial position of the cloud in the EW direction. We find a shift of -2079(31) kHz/mm, which corresponds to a mean velocity of 1.15(2) (mm/s)/mm. This corresponds to a mean Coriolis shift of 12.6(2) $\mu$Gal/mm, in reasonable agreement with the direct gravimeter measurement of 14.2(1.1) $\mu$Gal/mm. We also measure the asymetry of the distribution by calculating its skewness, defined as $1/\sigma^3\int f(v)(v-\bar{v})^3 dv$, where $\sigma$ is the rms velocity and $f(v)$ the normalized velocity distribution. We find a skewness of -0.16(1)/mm.
                                        
\begin{figure}[ht]
        \centering
        \includegraphics[width=8 cm]{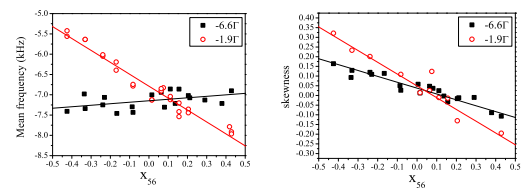}
\caption{Mean frequency and skewness of the effective distribution as a function of the power unbalance in the EW direction. The mean frequency is the mean of the Raman frequency spectrum, corrected from the hyperfine transition frequency.}
        \label{fig:x56}
   \end{figure} 
 
\section{Mean velocity versus molasses unbalance} 
 
We then measure the mean velocity as a function of the intensity unbalance. The results are presented on figure \ref{fig:x56}. We find -1.7(1) (mm/s)/unit of $x_{56}$ for a detuning of $-1.9\Gamma$ and 0.2(1) (mm/s)/unit of $x_{56}$ for $-6.6\Gamma$. This corresponds to Coriolis shifts of $18.6\mu$Gal/unit of $x_{56}$ at $-1.9\Gamma$ and $-2.2\mu$Gal/unit of $x_{56}$ at $-6.6\Gamma$. We observe a large reduction of the effect at the larger detuning, in agreement with the behaviour observed above on the gravity measurements. The effect of the Coriolis acceleration at $-1.9\Gamma$ is smaller here than measured previously in \cite{Louchet-Chauvet2011}. We attribute this difference to the lower intensity we have here in the molasses beams (1 mW/cm$^2$ instead of 1.6 mW/cm$^2$ for the measurements of \cite{Louchet-Chauvet2011}).
It is also noticeable that, though the scatter in the data at $-6.6\Gamma$, of order of 500 Hz ptp, is relatively large with respect to the trend, there seems to be resolved offset Doppler shifts of a few hundreds of Hz (with respect to the recoil shift of -7.5 kHz), which corresponds to mean velocities pointing towards the East, and thus negative Coriolis shifts of a few $\mu$Gal. This indicates that though the trend is reduced, the mean Coriolis shift is not null.


As for the skewness, they are found to differ by a factor of 2: -0.6/unit of $x_{56}$ at $-1.9\Gamma$ and -0.3/unit of $x_{56}$ at $-6.6\Gamma$. The skewness when expressed as a function of the displacements are close: -0.12/mm at $-1.9\Gamma$ and -0.11/mm at $-6.6\Gamma$, and slightly smaller than the skewness as a function of the initial position previously measured. This result tends to indicate that the skewness depends not only on the position but also on an asymetry of the {\it real} distribution due to the unbalance of the molasses beams. Interestingly, the skewness is null for $x_{56}=0.08$ at $-1.9\Gamma$ and $x_{56}=0.125$ at $-6.6\Gamma$. If we attribute the skewness to the effect of the initial position only, this indicates that the atoms do not fall at the center of the detection when the unbalance is null, but that we have to shift their initial position towards the East direction in order to find a symetric distribution. The corresponding displacements are found to be of $0.08\times 4.95\simeq0.4$ mm at $-1.9\Gamma$ (and $0.125\times 2.67\simeq0.33$ mm at $-6.6\Gamma$). This displacement could be due to a shift of the initial position with respect to the center of the MOT chamber, or more probably to a residual tilt of the vacuum chamber with respect to verticality. Indeed, the tilt of the experiment is set in order for the Raman retroreflecting mirror to be perfectly horizontal (to within about 10 microradians), but we cannot guarantee that this mirror is sitting in its support perfectly perpendicular to the "vertical" mechanical axis of the vaccum chamber. Considering the distance between the MOT and the detection of 20 cm, this displacement of about 0.4 mm corresponds to a tilt of the mirror with respect to the axis of the chamber of only 2 mrad. Under these conditions, we expect that when the beams are perfectly balanced, the atoms falls off from the center by -0.4 mm: the Coriolis shift due to this position offset is expected to be $14.2\times-0.4\simeq-5.7\mu$Gal. This shift could thus in principle be canceled by shifting the initial position of the atoms by the opposite amount, using for instance additional coils to displace the center of the MOT.

\begin{figure}[ht]
        \centering
      \includegraphics[width=8 cm]{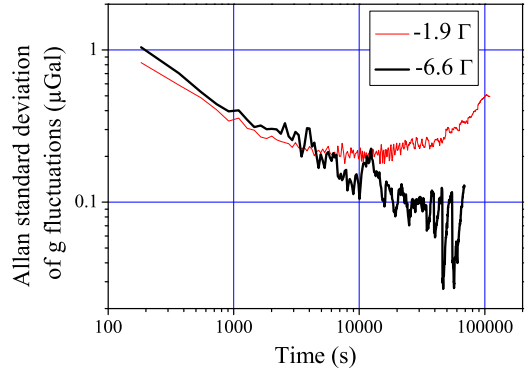}
\caption{Allan standard deviations of gravity fluctuations, calculated from the mean values of the effective velocity distributions, at detunings $-1.9\Gamma$ (thin line) and $-6.6\Gamma$ (thick line).}
        \label{fig:runscans}
   \end{figure}

In order to estimate the impact of the molasses detuning on the stability of the gravity measurement, we have performed consecutive measurements of the velocity distribution over a few days, for the two above detunings, and for molasses beams balanced at the beginning of the measurements. Figure \ref{fig:runscans} displays the Allan standard deviations of the gravity shifts calculated from the mean velocities of the spectra. The short term sensitivity, which corresponds to about 1 $\mu$Gal at 200 s, is limited by the noise on the velocity spectrum, which is dominated by atom number fluctuations. We observe a white noise averaging up to about 3000 s at a detuning of $-1.9\Gamma$. The stability then flickers at the level of $0.2 \mu$Gal then increases due to the drift of the mean velocity. The stability is better at a detuning of $-6.6\Gamma$: it averages down to better than $0.1 \mu$Gal after 20000s.   
 
\begin{figure}[ht]
        \centering
        \includegraphics[width=8 cm]{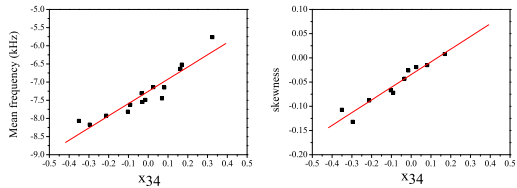}
\caption{Mean frequency and skewness of the effective distribution as a function of the power unbalance in the NS direction.}
        \label{fig:x34}
   \end{figure}

Because of the homogeneity of the detection system in the NS axis of the collection optics (better than 1\% over 2 cm \cite{Louchet-Chauvet2011}), the convolution with the detection doesn't affect the velocity distribution along the NS direction. The velocity distribution along this axis is thus expected to be the same as the real velocity distribution of the cloud. A fit with a Lorentzian b gives $b=1.9$ and $v_c=18.4$ mm/s, significantly different from the parameters of the EW distribution. Measurements of the mean velocity and skewness as a function of the intensity unbalance along the axis 34 (NS) are displayed in figure \ref{fig:x34}. The detuning is chosen here to be $-6.6\Gamma$. Linear fits to the data give a mean velocity of 1.8 mm/s/unit of $x_{34}$ and a skewness of 0.26/unit of $x_{34}$. In the present case, the mean velocity is directed towards the weakest beam of the molasses, as expected. The non-zero skewness of the distribution also shows that the effect of the unbalance is not only a change of the mean velocity but also an asymetry of the velocity distribution. The displacement is measured to be 3.61(8) mm/unit of $x_{34}$, larger than in the EW direction. The difference is due to the weaker magnetic field gradient of the MOT in the NS direction, and to a different cooling laser power in the NS beams. Indeed, as the atoms are captured from the intense beam of the 2D MOT aligned in the NS axis, we use twice larger intensities in the 1-2-3-4 beams than in the EW beams. 

 
\section{Gravity measurements with velocity selected atoms}

The possibility to velocity select the atoms in the horizontal direction allows to make a direct measurement of the interferometer phase shift as a function of the selected velocity. The velocity selection was performed using a sequence that prepares atoms in the $|F=1, m_F=0\rangle$ state in a narrow velocity distribution (whose FWHM is on the order of 0.6$v_r$) centered on a controlled mean velocity. The preparation sequence starts as above. After the "double-kick" Raman pulse sequence (see Annex), a last microwave pulse exchanges the populations of the $|F=2, m_F=0\rangle$ and $|F=1, m_F=0\rangle$, so that the atoms doubly selected are now in the $|F=1, m_F=0\rangle$. A final pusher pulse removes the atoms in $|F=2, m_F=0\rangle$. This selection scheme thus uses three microwave (MW) pulses, three pusher (P) pulses and two horizontal Raman pulses (HR) with the following sequence: MW+P+HR+MW+P+HR+MW+P.

\begin{figure}[ht]
        \centering
        \includegraphics[width=8 cm]{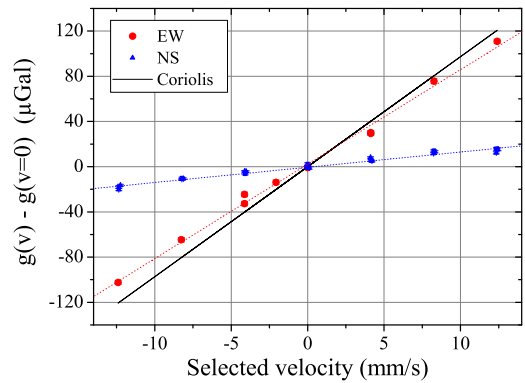}
\caption{Gravity measurements with velocity selected atoms in the NS and EW directions, and the expected shift due to Coriolis acceleration.}
        \label{fig:coriolis}
   \end{figure}

We performed differential measurements between two different selected velocities: the first measurement is a reference measurement for which the selected velocity is zero, the second corresponds to a non-zero selected velocity. Figure \ref{fig:coriolis} displays the results of the difference between these two measurements (the second measurement minus the first) for different selected velocities along the EW and NS directions, ranging from -12 to 12 mm/s. Here, the velocity is positive when pointing towards the West, so that the Coriolis shift is expected to be positive. We observe as expected a linear behaviour along the EW direction, of  $8.3(2)\mu$Gal/(mm/s), but also along the NS direction, of $1.3(1)\mu$Gal/(mm/s). We interpret the linear trend along the NS direction as due a residual Coriolis effect due to the imperfect orientation of the experiment, and thus of the $k_{eff}$ horizontal Raman wavevectors, with respect to the real NS and EW directions. This corresponds to an error of about 9$^{\circ}$ on the orientation of the experiment. We have carefully measured the orientation of the experiment and found a tilt of 8(1)$^{\circ}$. Combining these two projections of the EW velocity, we calculate a net Coriolis effect of $8.4(2) \mu$Gal/(mm/s), which is slightly weaker than the expected Coriolis bias of $9.72\mu$Gal/(mm/s). This difference could be due to the effect of wavefront aberrations, as atoms with non zero transverse velocities move in the Raman beams and might experience different phases at the three pulses. As for the impact of the clipping in the detection, it is expected to be smaller here than for previous measurements because the velocity distribution is much narrower. It is calculated to induce a correction of only 2.5\%. 

\begin{figure}[ht]
        \centering
        \includegraphics[width=8 cm]{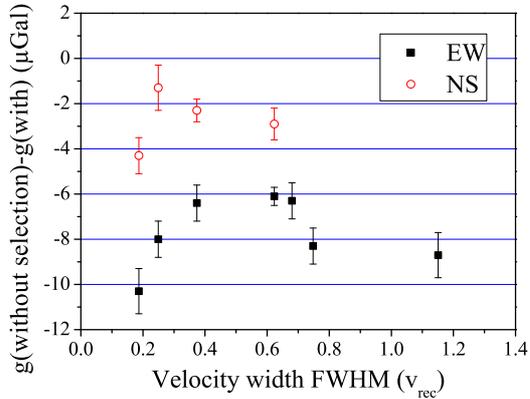}
\caption{Gravity measurements with velocity selected atoms in the NS and EW directions. Data points represent the difference between the case where the atoms are velocity selected using the double kick selection sequence and the case where there is no velocity selection, as a function of the width of the selected velocity distribution.}
        \label{fig:gselhor}
   \end{figure}

Finally, we measured the difference between the gravity values with and without horizontal selection. The measurement without horizontal selection was performed without vertical Raman selection either, using only a microwave selection to prepare the atoms in the $|F=1, m_F=0\rangle$ state. The measurement with horizontal selection was performed with a "double-kick" selection around the zero transverse velocity, in a narrow velocity distribution of FWHM ranging from 0.2 to $1.2v_{rec}$. The measurements were performed for a value of $x_{56}=0.06$ and at a detuning of $-6.6\Gamma$. The results of these differential measurements (differences between the unselected and selected case) are displayed in figure \ref{fig:gselhor}. We don't observe a net dependance on the width of the selected velocity distribution, but significant shifts of -7.0 $\mu$Gal on average in the case of the EW selection and of -2.7 $\mu$Gal in the case of the NS selection. We expect here that the Coriolis shift is null in the case of a horizontal velocity selection along the EW direction, because 1) the mean of the velocity distribution is null and 2) the effect of the clipping is drastically reduced when the width of the velocity distribution gets much narrower. As a consequence, the measured difference should correspond to the Coriolis shift in the case there is no selection. The shift of $-7\mu$Gal we find is in reasonable agreement with the estimate derived from the analysis of the skewness above, of $-5.7\mu$Gal. The difference between these two determinations can be attributed to changes in the bias due to wavefront aberrations. This effect could also explain the difference we measure when performing the velocity selection along the NS direction. Generally speaking, the contribution of this effect should depend on the width of the velocity distribution (extrapolating down to zero in the limit of zero temperature). This could explain the variations we observe in the measurements in figure \ref{fig:gselhor}. However, these variations appear hardly resolved here, considering the (1 sigma) statistical uncertainties in the measurements of order of 1 $\mu$Gal). 

\section{Conclusion}
We reported here an analysis of the Coriolis shift in an atom gravimeter. We show that the convolution with the response of the detection affects the averaging of this effect over the velocity distribution. In particular, the clipping due to the finite size of the detection field of view induces a dependance on the initial position of the atomic cloud. Remarkably, we find that the effect of this clipping almost compensates the effect due to a velocity drift along the EW direction, for an appropriate set of parameters in the preparation of the cold atom cloud. This compensation minimizes the sensitivity of the gravity measurement to the unbalance of the trapping laser beams and we have demontrated that it should improve the long term stability of the gravity measurement. We also perform interferometer measurements with atoms selected in a narrow velocity distribution. We find a reasonable agreement with the expected dependance on the initial mean velocity of the effect of Coriolis acceleration. Finally, we show that such a velocity selection can be used to reduce the shifts related to the transverse motion of the atoms in the Raman beam (Coriolis acceleration and wavefront aberrations) with respect to the situation where no selection is applied. Performing such a velocity selection in 2D would in principle allow an efficient suppression of these effects, at the expense of a complex sequence for the preparation of the atoms and a drastic reduction in atom number. Our study highligths the importance of the response of the detection in cold atom inertial sensors.

\section{Acknowlegments}
This research is carried on within the e-Mass and kNOW projects, which acknowledges the financial support of the EMRP. The EMRP was jointly funded by the European Metrology Research Programme (EMRP) participating countries within the European Association of National Metrology Institutes (EURAMET) and the European Union. B. C. thanks the Labex First-TF for financial support.

\section{Annex}
We describe in this annex our protocol to determine the effective velocity distribution, minimizing the perturbations related to the Raman selection process we use for its measurement, which are due to potential additional momentum transfer, light shifts and imperfect alignments.    
We start by measuring the horizontal velocity distribution by performing, instead of the standard vertical selection, a single horizontal Raman pulse after the microwave preparation. This Raman transition couples the $|F=1, m_F=0\rangle$ state, with initial momentum $|p\rangle$, to the state $|F=2, m_F=0,p+\hbar k_{eff}\rangle$, where $k_{eff}=k_1-k_2$ is the effective Raman wavevector, $k_1$ (resp $k_2$) being the wavector of the laser detuned from the $F=1\rightarrow F'$ (resp. $F=2\rightarrow F'$) transitions. When placing the two Raman beams such as diplayed as $k_1$ and $k_2$ in figure \ref{fig:setup}, $k_{eff}$ is pointing towards the East. If we choose $k_1'$ and $k_2$, $k_{eff}$ is pointing towards the South. The polarizations of the Raman beams are linear and horizontal (and thus perpendicular), with a quantization magnetic field being vertical. The Raman resonance condition is given by: $\omega_R=\omega_1-\omega_2=\omega_{HFS}+\omega_{rec}+\omega_{D}$. $\omega_{i}$ is the pulsation of the Raman laser i, $\omega_{HFS}$ is the pulsation of the hyperfine transition. $\omega_{rec}$ is the recoil term given by $\omega_{rec}=\hbar k_{eff}^2/2m$. For Raman beams crossing at $90^{\circ}$, $\omega_{rec}/2\pi\simeq7.5$kHz. $\omega_{D}$ is the Doppler term: $\omega_{D}=k_{eff}p$. This last term makes the Raman transition velocity selective and thus allows for a spectroscopic measurement of the velocity distribution.

\begin{figure}[ht]
        \centering
        \includegraphics[width=8 cm]{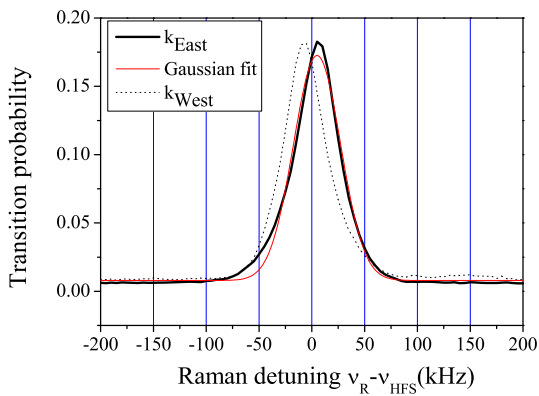}
\caption{Transition probability versus the detuning between the horizontal Raman lasers. Thick line: Raman spectrum with a momentum kick directed towards the East, thin line: a gaussian fit to the $k_{East}$ spectrum. Dotted line: momentum kick towards the West.}
        \label{fig:distribkdirect}
   \end{figure}

The velocity distribution is thus measured by scanning the Raman frequency difference $\nu_R$ across the hyperfine transition frequency $\nu_{HFS}$ and recording the transition probability. The Rabi frequency of about 5 kHz is about ten times smaller than the width of the atoms velocity distribution (expressed in terms of Doppler shifts : 1.8 kHz/mm/s), which ensures negligible contribution from the convolution with the Rabi excitation profile. Figure \ref{fig:distribkdirect} displays the measured spectrum. The comparison with a Gaussian fit clearly reveals a pronounced asymetry, which we attribute to the effect of the detection. Indeed, the selected atoms receive a momentum kick $\hbar k_{eff}$. This disymetrizes the distribution, due to clipping in the detection. This is illustrated in figure \ref{fig:clipping}, where atoms with large enough initial velocity in the direction of the momentum kick (and thus corresponding to positive Doppler shifts) are not detected, as they lie outside the field of view of the optical system that collects the fluorescence. This interpretation is confirmed by a second measurement, where prior to the Raman pulse, the atoms are transferred in the $|F=2,m_F=0\rangle$ state with an additional microwave $\pi$ pulse. The Raman beams then couple $|F=2, m_F=0,p\rangle$ to $|F=1, m_F=0,p-\hbar k_{eff}\rangle$, the resonance condition being now given by $\omega_R=\omega_1-\omega_2=\omega_{HFS}-\omega_{rec}+\omega_{D}$. The sign of the recoil shift is reversed, as well as the direction of the momentum kick. As expected, we observe a reversal of the asymetry on the spectrum (see the profile as a dotted line in figure \ref{fig:distribkdirect}).   

To prevent the horizontal Raman velocity selection from modifying the velocity distribution of the atoms, we use a different selection sequence. Before the Raman pulse, the atoms are prepared in the $|F=2, m_F=0\rangle$ state. A first horizontal Raman pulse transfers a narrow velocity class in the $|F=1,m_F=0\rangle$, the remaining atoms are then pushed away thanks to a pulse of light tuned on the $|F=2\rightarrow F'=3\rangle$ cycling transition. A second Raman pulse retransfers the selected atoms back into the $|F=2,m_F=0\rangle$ with an opposite momentum transfer. The net momentum transfer in the sequence is thus zero. The velocity distribution is finally obtained by scanning the frequency difference between the Raman lasers and measuring the number of atoms in the $|F=2,m_F=0\rangle$ after this "double-kick" selection. The resonance frequency is corrected from eventual light shifts from the Raman lasers. This light shift is measured by microwave spectroscopy, by measuring the microwave resonance frequency in the presence of Raman lasers set out of resonance. In addition, the direction of the wavevector is precisely aligned in the horizontal plane. Indeed, any misalignement will lead to a shift of the mean resonance frequency due to a vertical Doppler shift. We optimize the alignment of the horizontal Raman beams by nulling the variation of the mean resonance condition with the delay of the Raman pulse with respect to the drop time, and thus with the vertical velocity. In practice, the quality of this alignement is limited by the noise in the velocity distribution measurement induced by fluctuations in the number of atoms and in the Raman light shift. We estimate the residual uncertainty on the contribution to the frequency spectrum of a residual vertical Doppler shift to be about 200 Hz. This translates into an uncertainty in the determination of the Coriolis shifts of $1.2\mu$Gal.

\end{document}